\documentclass[prb,preprint]{revtex4-1}

\usepackage{amsmath}  
\usepackage{amsfonts} 
\usepackage{graphicx} 

\begin{document}

\title{Electrostatic responses of anisotropic dielectric films} 

\author{Hai-Yao Deng}
\email{haiyao.deng@gmail.com}
\affiliation{National Graphene Institute, University of Manchester, Booth St E, Manchester M13 9PL, United Kingdom}

\begin{abstract} 
We study the electrostatic responses (i.e. retardation effects due to the propagation of electromagnetic waves are ignored) of a linear homogeneous and anisotropic (LHA) dielectric film to an arbitrary external electrostatic potential, which may be generated by charges located either inside or outside it. A formalism is developed to calculate the polarization charges induced in the film. In our derivation, the idea is exploited that a physical boundary can be looked upon as a region of rapid variation in polarization rather than a simple geometric separation. With this no boundary conditions are needed in solving the relevant electrostatics problem. Our approach makes it clear that the responses consist of two contributions, one arising from the very presence of surfaces while the other existing even in an infinite medium. The approach can be applied not just to electrostatic but also many problems and is of great pedagogical value. In light of the results, we discuss graphene plasma waves under the influence of a LHA dielectric film such as a few-layer hexagonal boron nitride. It is found that the dispersion of these waves is strongly affected by the anisotropy at wavelengths comparable to the film thickness. 
\end{abstract}

\maketitle

\section{introduction}
\label{sec:1}
As well known in electrostatics, an exterior probe charge cannot induce volume polarization charges in a linear homogeneous and isotropic (LHI) dielectric, as the divergence of the displacement field $\mathbf{D}$ and hence that of the electric field $\mathbf{E}$ -- which is proportional to $\mathbf{D}$ in a LHI -- vanish in the body. Polarization charges appear only where inhomogeneity exists, e.g. at the interfaces and surfaces adjoining two LHI media. In the traditional textbook approach of evaluating the amount of polarization charges on the surface of -- for instance -- a semi-infinite medium (SIM), the electrostatic fields residing on the opposite sides of the surface are treated separately on a piece-wise homogeneity basis and then joined by a set of boundary conditions requiring the continuity of the normal component of $\mathbf{D}$ and that of the electrostatic potential $\Phi$ at the surface.~\cite{landauEM} In this approach, a surface is more of a geometric separation rather than a physical entity. 

The purpose of the present work is two-fold. First, we exploit an alternative approach for studying the electrostatic responses of dielectrics. In this new approach, an interface is treated, as in an atomistic approach, as a physical region over which electric polarization undergoes rapid variation. Though atomistically unknown \textit{a priori}, such variations can be fixed on a macroscopic scale by which the interfacial region appears infinitely thin. As a result, boundary conditions are done away with in totality. It should be mentioned that this macroscopic description of surfaces has recently been employed in the study of surface plasma waves,~\cite{deng2017,deng2019,deng2018} i.e. electron density ripples propagating on the surface of conductors. Similar ideas have been used in a very different context, i.e. the study of plasma waves in bounded two-dimensional electron gas.~\cite{fetter} 

Second, we apply the approach to study the responses of linear homogeneous anisotropic (LHA) dielectric films. While the electrostatics of LHI media make a standard part of any textbook on electromagnetism, LHA media have been unduly less exposed despite the fact that the latter constitute a big portion in real life. Existing work on LHA dielectrics have been focused on SIMs and often in a geophysical context.~\cite{li1997,mele2001} However, in the world of nanometer electronics, one has to deal with LHA dielectrics of nanometric thickness. It is therefore desirable to conduct a systematic exposition of the electrostatic responses of LHA films and clarify the thickness effects. Differing from a LHI dielectric, a LHA dielectric can host both volume and boundary polarization charges, which can vary significantly depending on the film thickness. We derive and solve a set of algebraic equations for determining the responses of a LHA film to an arbitrary external potential, which might be generated by probe charges located inside or outside the film. With the results we examine the polarization charges induced in a film by a point charge. Two cases are considered, corresponding to the point charge being located inside and outside the film, respectively. As an additional example, we employ the results to discuss the behaviors of plasma waves in a mono-layer graphene under the influence of a LHA film such as a few-layer hexagonal boron nitride. These materials belong to the family of the so-called van der Waals materials, which are a hot topic in contemporary research.~\cite{vdW} We find that the plasma wave dispersion is strongly affected by the anisotropy at wavelengths comparable to the film thickness. 

\section{Macroscopic description of an interface}
\label{sec:2}
\begin{figure*}
\begin{center}
\includegraphics[width=0.95\textwidth]{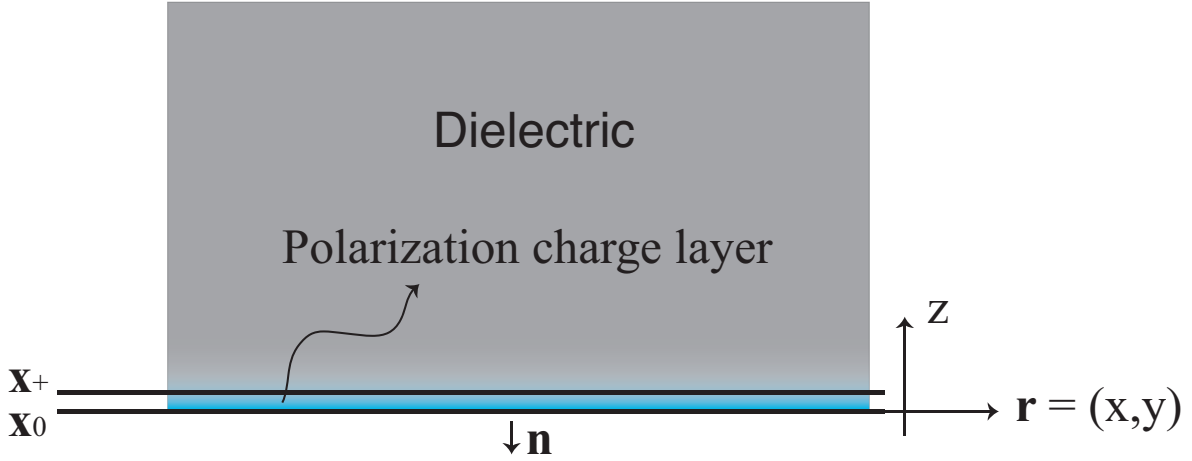}
\end{center}
\caption{Sketch of the polarization charge layer on the surface of a semi-infinite dielectric. Even though the layer has an infinitesimal thickness on the macroscopic scale, one should distinguish between the two planes comprised of points $\mathbf{x}_+ = (\mathbf{r},0_+)$ and $\mathbf{x}_0 = (\mathbf{r},0)$, respectively. Here $0_+$ denotes the positive infinitesimal. The surface consists of points $\mathbf{x}_0$, and $\mathbf{n}$ is the direction vector of its outward normal. One should note that, in the celebrated  expression for the areal density of the polarization charges, namely $\mathbf{P}\cdot\mathbf{n}$, $\mathbf{P}$ is the polarization at $\mathbf{x}_+$ rather than $\mathbf{x}_0$, as is obvious in the boundary conditions used in the traditional approach to the electrostatics of dielectrics. \label{fig:1}}
\end{figure*}

From an atomistic point of view, an interface is not just a geometric separation but a physical region in which the polarization rapidly evolves from one side to the other. For the sake of illustration, let us imagine bringing in touch two semi-infinite dielectrics, A and B, and an atomically thin interface forms amid them. We shall assume that the interface is macroscopically flat and extends within the $x-y$ plane so that its normal points along $z$-direction, see Fig.~\ref{fig:1}. The polarization in the entire space is denoted by $\mathbf{p}(\mathbf{x})$, where $\mathbf{x} = (\mathbf{r},z)$ denotes a point in the space with $\mathbf{r} = (x,y)$ being the planar projection. Quite generally, far away from the interface, we expect that $\mathbf{p}(\mathbf{x})$ takes on the bulk value $\mathbf{P}_{A/B}(\mathbf{x})$ on the A/B side. As one moves across the interface from A to B, $\mathbf{p}$ shall evolve from $\mathbf{P}_A$ to $\mathbf{P}_B$. One may introduce an evolution function $w(z)$, which approaches unity for $z$ lying deeply in A whereas it approaches zero for $z$ lying deeply in B. Then, $\mathbf{p}(\mathbf{x}) = w(z)\mathbf{P}_A(\mathbf{x}) + (1-w(z))\mathbf{P}_B(\mathbf{x})$. The exact form of $w(z)$ in the interfacial region depends on atomistic details. Nevertheless, on the macroscopic scale by which the interfacial region appears infinitely thin, one can approximate $w(z)$ by the Heaviside step function $\Theta(z)$ regardless of the atomistic details, presuming that the surface be located at $z=0$ without loss of generality. As such, the macroscopic description of an interface is fixed in a fairly generic manner.~\cite{deng2019} In the case of a SIM, we have B as the vacuum and A as the medium. With $\mathbf{P}_B \equiv 0$ and writing $\mathbf{P}_A$ simply as $\mathbf{P}$, we can write
\begin{equation}
\mathbf{p}(\mathbf{x}) = \Theta(z) \mathbf{P}(\mathbf{x}). \label{p}
\end{equation}
Note that equation (\ref{p}) gives the polarization in the entire system (the dielectric plus the vacuum) and hence no boundary is needed in this global view. The polarization charge density is then obtained as
\begin{equation}
\rho(\mathbf{x}) = -\partial_\mathbf{x}\cdot\mathbf{p}(\mathbf{x}) = \rho_b(\mathbf{x}) + \rho_s(\mathbf{x}), \label{rho}
\end{equation} 
where $\rho_b(\mathbf{x}) = -\partial_\mathbf{x}\cdot\mathbf{P}(\mathbf{x})$ is the density of volume charges in the dielectric and $\rho_s(\mathbf{x}) = -\Theta'(z) P_z(\mathbf{x}_0)$ gives the density of charges existing on the surface. Here $\Theta'(z) = \partial_z \Theta(z)$ and $\mathbf{x}_0 = (\mathbf{r},0)$ represents a point on the surface. 

As an illustration of the usefulness of Eq.~(\ref{rho}), let us consider a LHI characterized by a constant susceptibility $\chi$, i.e. $\mathbf{P}(\mathbf{x}) = \chi \mathbf{E}(\mathbf{x})$. Assuming no interior probe charges, i.e. $\rho_{ext}(\mathbf{x}) = 0$ for $\mathbf{x}$ lying inside the medium and hence $\partial_\mathbf{x}\cdot\mathbf{E}(\mathbf{x}) = 4\pi (\rho_{ext}(\mathbf{x}) + \rho(\mathbf{x})) = 4\pi \rho(\mathbf{x})$, we find from Eq.~(\ref{rho}) that $\rho(\mathbf{x}) = - \Theta'(z) (\chi/\epsilon) E_z(\mathbf{x}_0)$, where $\epsilon = 1 + 4\pi \chi$ is the dielectric constant. As expected, $\rho(\mathbf{x})$ is concentrated on the surface. To bring the expression into the familiar form, let us note that $E_z(\mathbf{x}_0)/\epsilon = E_z(\mathbf{x}_+)$, where $\mathbf{x}_+ = (\mathbf{r},0_+)$ with $0_+$ being the positive infinitesimal (see Fig.~\ref{fig:1} for illustration). See that $\mathbf{x}_0$ and $\mathbf{x}_+$ represent points on the planes sandwiching the polarization charge layer, as shown in Fig.~\ref{fig:1}. Now we obtain $\rho(\mathbf{x}) = - \Theta'(z)P_z(\mathbf{x}_+)$, which implies an areal density of $\mathbf{P}(\mathbf{x}_+)\cdot \mathbf{n}$, where $\mathbf{n}$ is the unit vector pointing along the outward normal of the surface, as expected of the traditional approach.

\section{Responses of LHA dielectric films}
\label{sec:3}
The dielectric film under consideration has two infinitely extended flat surfaces at $z=0$ and $z=L>0$, respectively. It is characterized by a constant susceptibility tensor of the following diagonal form, $$\chi = \text{diag}(\chi_\parallel,\chi_\parallel,\chi_\perp).$$ This is not the most general form but captures the properties of a large variety of materials including van der Waals materials. It shall prove convenient to introduce the dielectric tensor components as $\epsilon_{\parallel} = 1 + 4\pi \chi_\parallel$ and $\epsilon_\perp = 1 + 4\pi \chi_\perp$, together with the anisotropy parameter $\gamma = \sqrt{\epsilon_\perp/\epsilon_\parallel}$. We can write $\epsilon_\parallel = \epsilon/\gamma$ and $\epsilon_\perp = \gamma \epsilon$, where $\epsilon = \sqrt{\epsilon_\parallel \epsilon_\perp}$.

In analogy with Eq.~(\ref{p}), the global polarization in this system can be written as
\begin{equation}
\mathbf{p}(\mathbf{x}) = \mathbf{P}(\mathbf{x})\left[\Theta(z) - \Theta(z-L)\right],
\end{equation}
where $P_i(\mathbf{x}) = \sum_j\chi_{ij} E_j(\mathbf{x})$ with $i,j=x,y,z$. The polarization charge density then reads
\begin{equation}
\rho(\mathbf{x}) = -\partial_\mathbf{x}\cdot \mathbf{p}(\mathbf{x}) = - \partial_\mathbf{x}\cdot\mathbf{P}(\mathbf{x}) - P_z(\mathbf{x})\left[\Theta'(z) - \Theta'(z-L)\right]. \label{4}
\end{equation}
Here the first term, which vanishes in a LHI but not in a LHA, gives the density of volume charges while the second one gives the density of surface charges.

We introduce the planar Fourier transform pair for a field quantity $f(\mathbf{x})$, 
$$f_\mathbf{k}(z) = \int d^2\mathbf{r} e^{-i\mathbf{k}\mathbf{r}} f(\mathbf{x}),\quad f(\mathbf{x}) = \int \frac{d^2\mathbf{k}}{4\pi^2} e^{i\mathbf{k}\mathbf{r}} f_\mathbf{k}(z), $$
where $\mathbf{k}$ is the wave vector along the surface. As our system is linear, it suffices to consider one Fourier component, say $\mathbf{k}$. Subsequently, we shall suppress the index $\mathbf{k}$ and write $f_\mathbf{k}(z)$ simply as $f(z)$, which should not be confused with $f(\mathbf{x})$ as their arguments are of different character. Equation (\ref{4}) is then rewritten
\begin{equation}
\rho(z) = -\nabla\cdot\mathbf{P}(z) - \left[P_z(0)\Theta'(z) -P_z(L)\Theta'(z-L)\right], \label{5}
\end{equation}
where $z$ is confined to the dielectric and $\nabla = (i\mathbf{k},\partial_z)$.

Our task is to calculate $\rho(z)$ induced by an external probe potential $\phi_{ext}(z)$, which is supposed to be generated by a charge of density $\rho_{ext}(z)$. The electrostatic potential due to $\rho(z)$ is denoted by $\phi(z)$. It is well known and can be easily shown that
\begin{equation}
\begin{pmatrix}
\phi_{ext}(z) \\
\phi(z)
\end{pmatrix}
=
\frac{2\pi}{k} \int dz' e^{-k|z-z'|} 
\begin{pmatrix}
\rho_{ext}(z') \\
\rho(z')
\end{pmatrix}. \label{6}
\end{equation}
Here $k = |\mathbf{k}|$ and the zero point of the potential has been set at infinity. 

\subsection{General formalism}
\label{sec:3.1}
Note that $-\nabla\cdot\mathbf{P}(z) = -4\pi\chi_\perp (\rho(z) + \rho_{ext}(z)) + k^2 \delta\chi \Phi(z)$, where $\delta\chi = \chi_\perp - \chi_\parallel$ and, as mentioned earlier, $\Phi(z) = \phi(z) + \phi_{ext}(z)$ is the total electrostatic potential. Substituting this in Eq.~(\ref{5}) yields
\begin{equation}
\rho(z) = \rho_b(z) + \rho_s(z), 
\end{equation}
where 
\begin{equation}
\rho_b(z) = \frac{k^2 \delta\chi}{\epsilon_\perp} \Phi(z) + \frac{1-\epsilon_\perp}{\epsilon_\perp} \rho_{ext}(z)\label{8}
\end{equation} 
is the charge density that exists also in an infinite medium and 
\begin{equation}
\rho_s(z) = \frac{\chi_\perp}{\epsilon_\perp} \left[E_z(L)\Theta'(z-L) - E_z(0)\Theta'(z)\right] \label{9}
\end{equation}
denotes the charge density purely due to the presence of surfaces. It should be clear that, Eqs.~(\ref{6}) - (\ref{9}) make a closed set and can be solved already to determine the polarization charges. Below we prescribe them in a more tractable way. 

To make progress, we extend $\rho(z)$ in the following cosine series,
\begin{equation}
\rho(z) = \sum^\infty_{n=0} \rho_n \cos(q_n z), \quad \rho_n = \frac{1}{L_n}\int^L_0 dz \cos(q_nz)\rho(z),
\end{equation}
where $q_n = \pi n/L$ and $L_n = L/(2-\delta_{n,0})$ with $\delta_{n,0}$ being zero unless $n=0$. It shall prove convenient to split $\rho(z)$ into a symmetric part $\rho_+(z)$ and an anti-symmetric part $\rho_-(z)$, where $\rho_{+}$ collects all the terms with even $n$ and $\rho_-$ those with odd $n$. Obviously, $\rho_+$ is symmetric about the mid-plane of the film while $\rho_-$ is anti-symmetric. In terms of $\rho_n$, one can recast
\begin{equation}
\phi(z) = \sum^\infty_0 \frac{2\pi \rho_n}{k^2+q^2_n}\left[2\cos(q_nz) - e^{-kz} - (-1)^n e^{-k(L-z)}\right] \label{11}
\end{equation}
for $z$ lying in the dielectric. It follows that $\phi'(z) = \partial_z\phi(z)$ takes on the following values at the surfaces,
\begin{equation}
\begin{pmatrix}
\phi'(0) \\
\phi'(L)
\end{pmatrix}
=
\sum^\infty_{n=0} \frac{2\pi k\rho_n}{k^2+q^2_n} 
\begin{pmatrix}
1 - (-1)^ne^{-kL} \\
(-1)^n \left[(-1)^ne^{-kL} - 1\right]
\end{pmatrix}. \label{12}
\end{equation}
Now we have $E_z(0/L) = -\phi'(0/L) - \phi'_{ext}(0/L)$. 

Let $f_n = L^{-1}_n\int^L_0 dz \cos(q_nz)f(z)$ for $f(z)$. We can write $\phi_n$ as a linear function of $\rho_n$, i.e. $\phi_n = \sum_{n'} M_{n,n'} \rho_{n'}$, where the matrix elements can be established from Eq.~(\ref{11}) as
\begin{equation}
M_{n,n'} = \frac{4\pi}{k^2+q^2_n}\left(\delta_{n,n'} - \frac{k}{k^2+q^2_{n'}}\frac{1-(-1)^ne^{-kL}}{2L_n}\left(1+(-1)^{n+n'}\right)\right). 
\end{equation}
From Eq.~(\ref{8}) we obtain
\begin{equation}
\rho_{b,n} = \frac{k^2\delta\chi}{\epsilon_\perp} \sum_{n'}M_{n,n'}\rho_{n'} + \frac{k^2\delta\chi}{\epsilon_\perp} \phi_{ext,n} + \frac{1-\epsilon_\perp}{\epsilon_\perp} \rho_{ext,n}. \label{14}
\end{equation}
Analogously, from Eq.~(\ref{9}) it follows that
\begin{equation}
\rho_{s,n} = \frac{\chi_\perp}{\epsilon_\perp}\frac{1}{L_n} \left[\sum_{n'}\frac{2\pi k\rho_{n'}}{k^2+q^2_{n'}}\left(1-(-1)^{n'}e^{-kL}\right)\left(1+(-1)^{n+n'}\right) + \phi'_{ext}(0) - (-1)^n\phi'_{ext}(L)\right]. \label{15}
\end{equation}
Adding Eqs.~(\ref{14}) and (\ref{15}), one ends up with a linear inhomogeneous algebraic equation for $\rho_n$, which can then be solved for any $\rho_{ext,n}$. 

It is interesting to note that, due to the factor $1+(-1)^{n+n'}$ appearing in both $M_{n,n'}$ and $\rho_{s,n}$, the symmetric sector $\rho_+(z)$ and the anti-symmetric sector $\rho_-(z)$ are actually decoupled. As such, we can deal with them separately. To this end, let us define $\rho^+_l = \rho_{2l}$ and $\rho^-_{l} = \rho_{2l+1}$ with $l = 0,1,...$ being an integer. It immediately follows that
\begin{equation}
\rho^+_l = \sum^\infty_{l' = 0} \mathcal{M}^+_{ll'}\rho^+_{l'} + \frac{k^2\delta\chi}{\epsilon_\perp} \phi_{ext,2l}+ \frac{1-\epsilon_\perp}{\epsilon_\perp} \rho_{ext,2l} + \frac{2-\delta_{l,0}}{L}\frac{\chi_\perp}{\epsilon_\perp}\left(\phi'_{ext}(0) - \phi'_{ext}(L)\right), \label{16}
\end{equation}
where the matrix $\mathcal{M}^+$ is given by
\begin{equation}
\mathcal{M}^+_{ll'} = \frac{k^2\delta\chi}{\epsilon_\perp}M_{2l,2l'} + \frac{\chi_\perp}{\epsilon_\perp}\frac{(2-\delta_{l,0})(1-e^{-kL})}{L} \frac{4\pi k}{k^2+q^2_{2l'}}. 
\end{equation}
Similarly, we have
\begin{equation}
\rho^-_l = \sum^\infty_{l' = 0} \mathcal{M}^-_{ll'}\rho^-_{l'} + \frac{k^2\delta\chi}{\epsilon_\perp} \phi_{ext,2l+1}+ \frac{1-\epsilon_\perp}{\epsilon_\perp} \rho_{ext,2l+1} + \frac{2}{L}\frac{\chi_\perp}{\epsilon_\perp}\left(\phi'_{ext}(0) + \phi'_{ext}(L)\right), \label{18}
\end{equation}
where the matrix $\mathcal{M}^-$ is given by
\begin{equation}
\mathcal{M}^-_{ll'} = \frac{k^2\delta\chi}{\epsilon_\perp}M_{2l+1,2l'+1} + \frac{\chi_\perp}{\epsilon_\perp}\frac{2(1+ e^{-kL})}{L} \frac{4\pi k}{k^2+q^2_{2l'+1}}. \label{19}
\end{equation}
These algebraic equations (\ref{16}) - (\ref{19}) fully determine the electrostatic responses of the film. 

The solutions to the above equations are readily to be found. Formally, they can be written as
\begin{equation}
\rho^+_l = \frac{\frac{\pi (2-\delta_{l,0})(1-e^{-kL})}{L}\left[(\epsilon_\parallel - 1) k^2 + (\epsilon_\perp - 1) q^2_{2l}\right]\bar{\rho}_+ + S^+_l}{\epsilon_\parallel k^2 + \epsilon_\perp q^2_{2l}}, \label{20}
\end{equation}
where $\bar{\rho}_+$ is independent of $l$ and $S^+_l = (k^2+q^2_{2l}) \tilde{S}^+_l$, given by
\begin{equation}
\bar{\rho}_+ = \frac{1}{\pi}\sum_{l} \frac{k \rho^+_l}{k^2 + q^2_{2l}}, \quad \tilde{S}^+_l = k^2 \delta\chi \phi_{ext,2l} + (1-\epsilon_\perp) \rho_{ext,2l} + \frac{2-\delta_{l,0}}{L}\chi_\perp(\phi'_{ext}(0) - \phi'_{ext}(L)). 
\end{equation}
Further manipulations show that
\begin{equation}
\bar{\rho}_+ = \frac{\frac{1}{\pi}\sum_l \frac{k}{k^2 + q^2_{2l}} \frac{S^+_l}{\epsilon_\parallel k^2 + \epsilon_\perp q^2_{2l}}}{1 - \sum_l \frac{k}{k^2 + q^2_{2l}}\frac{(2-\delta_{l,0})(1-e^{-kL})}{L} \frac{(\epsilon_\parallel - 1) k^2 + (\epsilon_\perp - 1) q^2_{2l}}{\epsilon_\parallel k^2 + \epsilon_\perp q^2_{2l}}}. \label{rhop}
\end{equation}
The expressions relevant for $\rho^-_l$ are analogous. For the sake of completeness, they are written down as follows
\begin{eqnarray}
&~& \rho^-_l = \frac{\frac{2\pi (1+ e^{-kL})}{L}\left[(\epsilon_\parallel - 1) k^2 + (\epsilon_\perp - 1) q^2_{2l+1}\right]\bar{\rho}_- + S^-_l}{\epsilon_\parallel k^2 + \epsilon_\perp q^2_{2l+1}}, \quad S^-_l = (k^2+q^2_{2l+1})\tilde{S}^-_l,\\
&~& \tilde{S}^-_l = k^2 \delta\chi \phi_{ext,2l+1} + (1-\epsilon_\perp) \rho_{ext,2l+1} + \frac{2}{L}\chi_\perp(\phi'_{ext}(0) + \phi'_{ext}(L)), \\
&~& \bar{\rho}_- = \frac{1}{\pi} \sum_{l} \frac{k \rho^-_l}{k^2 + q^2_{2l+1}} = \frac{\frac{1}{\pi}\sum_l \frac{k}{k^2 + q^2_{2l+1}} \frac{S^-_l}{\epsilon_\parallel k^2 + \epsilon_\perp q^2_{2l+1}}}{1 - \sum_l \frac{k}{k^2 + q^2_{2l+1}}\frac{2 (1+ e^{-kL})}{L} \frac{(\epsilon_\parallel - 1) k^2 + (\epsilon_\perp - 1) q^2_{2l+1}}{\epsilon_\parallel k^2 + \epsilon_\perp q^2_{2l+1}}}. \label{200}
\end{eqnarray}
Equations.~(\ref{rhop}) and (\ref{200}) represent the key results of this paper, which can be used to calculate the polarization charges for any given $\phi_{ext}$. The expressions of $\rho^\pm_l$ imply that there are two general contributions to the polarization charges.~\cite{deng2018} One of these contains $\bar{\rho}_\pm$. This contribution exists only in the presence of the surfaces. The other is directly due to $S^\pm_l$ and exists even if there is no surface. 

We point out that $\phi_{ext,n}$ and $\phi'_{ext}(0/L)$ can be written in terms of $\rho_{ext,n}$. Actually, $\phi_{ext,n} = \sum_{n'}M_{n,n'}\rho_{ext.n'}$ and $\phi'_{ext}(0/L)$ can be obtained from Eq.~(\ref{12}) with $\rho_{ext,n}$ in place of $\rho_n$. It is worth noting that
\begin{equation}
\phi'_{ext}(0) \pm \phi'_{ext}(L) = \sum^\infty_{n=0} \frac{2\pi k\rho_{ext,n}}{k^2+q^2_n} \left(1-(-1)^ne^{-kL}\right)\left(1\mp (-1)^n\right).
\end{equation}
More precisely, $\phi_{ext,2l} = \sum_{l'}M_{2l,2l'}\rho_{ext,2l'}$, $\phi_{ext,2l+1} = \sum_{l'}M_{2l+1,2l'+1}\rho_{ext,2l'+1}$ and $\phi'_{ext}(0)-\phi'_{ext}(L)$ only involves $\rho_{ext,2l}$ whereas $\phi'_{ext}(0)+\phi'_{ext}(L)$ only involves $\rho_{ext,2l+1}$. These properties ensure that a (anti-) symmetric $\rho_{ext}$ only induces a (anti-) symmetric $\rho$, in accord with the symmetry of the system. 

\subsection{The limit $kL\ll1$}
\label{sec:3.2}
Suppose $L$ is very small so that $\phi_{ext}(z)$ does not vary much across the film and $kL\ll1$. Then one may retain only the $n=0$ term for $\rho_b$, i.e. $\rho_b(z) \approx \rho_{b,0}$. The total volume charges then amounts to $\bar{\rho}_b = \int dz \rho_b(z) \approx L\rho_{b,0}$. This approximation, however, does not apply to $\rho_s$, which is localized and must have all terms present. It is thus useful to construct a separate formalism for this special case. By Eq.~(\ref{8}), we arrive at
\begin{equation}
\rho_{b,0} = \frac{k^2\delta\chi}{\epsilon_\perp}(\phi_0 + \phi_{ext,0}) + \frac{1-\epsilon_\perp}{\epsilon_\perp} \rho_{ext,0}, \label{21}
\end{equation}
From Eq.~(\ref{6}) one can show that 
\begin{equation}
\phi_0 \approx \frac{2\pi L}{k}\rho_{b,0} + \frac{2\pi}{k}\frac{1-e^{-kL}}{kL}(\rho_{s0}+\rho_{sL}),
\end{equation}
where $\rho_{s0} = -(\chi_\perp/\epsilon_\perp)E_z(0)$ and $\rho_{sL} = (\chi_\perp/\epsilon_\perp)E_z(L)$ are the areal density of polarization charges localized on the surface $z=0$ and $z=L$, respectively. Substituting this expression in Eq.~(\ref{21}) leads to
\begin{equation}
\left(\frac{\epsilon_\perp}{2\pi kL\delta\chi}-1\right) \bar{\rho}_b - \frac{1-e^{-kL}}{kL}(\rho_{s0} + \rho_{sL}) = \frac{k}{2\pi}\left(\phi_{ext,0} + \frac{1-\epsilon_\perp}{k^2\delta\chi}\rho_{ext,0}\right). \label{23}
\end{equation}
On using the following relations
\begin{equation}
\begin{pmatrix}
\phi'(0)/2\pi \\
\phi'(L)/2\pi
\end{pmatrix}
=
\bar{\rho}_b
\begin{pmatrix}
1 \\
-1
\end{pmatrix}
+
\begin{pmatrix}
\rho_{s0} + \rho_{sL}e^{-kL} \\
- \rho_{sL} - \rho_{s0}e^{-kL}
\end{pmatrix},
\end{equation}
we obtain
\begin{eqnarray}
\label{25} &~& -\bar{\rho}_b + \beta^{-1} \rho_{s0} - e^{-kL}\rho_{sL} = \frac{\phi'_{ext}(0)}{2\pi}, \\
\label{26} &~& \bar{\rho}_b + e^{-kL} \rho_{s0} - \beta^{-1} \rho_{sL} = \frac{\phi'_{ext}(L)}{2\pi},
\end{eqnarray}
where $\beta = (\epsilon_\perp - 1)/(\epsilon_\perp + 1)$. Now Eqs.~(\ref{23}), (\ref{25}) and (\ref{26}) make a closed set of algebraic equations and can be solved to get $\bar{\rho}_b, \rho_{s0}$ and $\rho_{sL}$. Clearly, only $\epsilon_\perp$ is involved.

Here are some general observations based on these equations. Firstly, by Eq.~(\ref{23}), we see that $$\bar{\rho}_b \approx L\rho_{ext,0} (1-\epsilon_\perp)/\epsilon_\perp + \frac{2\pi \delta\chi}{\epsilon_\perp}(\rho_{s0} +\rho_{sL}) kL + ...,$$ where "$...$" represents higher-order terms in $kL$. This means that the total volume charges are proportional to the total probe charges residing in the film to the lowest order in $kL$. If all probe charges are located outside, volume charges can be neglected for thin films, in which circumstances one obtains from Eqs.~(\ref{25}) and (\ref{26}) that 
$$\rho_{s0} \approx \frac{\beta}{2\pi}\frac{\phi'_{ext}(0) - \beta\phi'_{ext}(L)}{1-\beta^2}, \quad \rho_{sL} \approx - \frac{\beta}{2\pi}\frac{\phi'_{ext}(L) - \beta\phi'_{ext}(0)}{1-\beta^2}.$$ Secondly, Eqs.~(\ref{25}) and (\ref{26}) suggest that 
$$\rho_{s0} - \rho_{sL} = \frac{1}{\beta^{-1} + e^{-kL}} \frac{\phi'_{ext}(0) + \phi'_{ext}(L)}{2\pi}.$$ If the probe charges are symmetrically distributed, i.e. $\phi'_{ext}(0) + \phi'_{ext}(L) = 0$, the amounts of charges on the surfaces are equal and of the same sign, as expected of the reflection symmetry and in line with the general formalism set out in Sec.~\ref{sec:3.1}. 

\subsection{The semi-infinite limit}
\label{sec:3.3}
For $L\rightarrow\infty$, $e^{-kL}\approx0$ for all $k$ and the two surfaces are decoupled. The responses then reduce to that of a SIM. Now one notes that $\mathcal{M}^+$ and $\mathcal{M}^-$ degenerate to $\mathcal{M}$, where
\begin{equation}
\mathcal{M}_{l,l'} = \frac{k^2 \delta\chi}{\epsilon_\perp}\frac{4\pi}{k^2+Q^2_{l}}\left(\delta_{l,l'} - \frac{2k/L}{k^2+Q^2_{l'}}\right) + \frac{\chi_\perp}{\epsilon_\perp}\frac{2}{L}\frac{4\pi k}{k^2+Q^2_{l'}}, \label{27}
\end{equation}
where $Q_l = q_{2l}$. Adding Eq.~(\ref{16}) to Eq.~(\ref{18}), we find 
\begin{equation}
\rho_l = \sum_{l'} \mathcal{M}_{ll'} \rho_{l'} + \frac{k^2 \delta\chi}{\epsilon_\perp}\phi_{ext,l} + \frac{1-\epsilon_\perp}{\epsilon_\perp} \rho_{ext,l} + \frac{4}{L} \frac{\chi_\perp}{\epsilon_\perp}\phi'_{ext}(0), \label{28}
\end{equation}
where we have defined
$$\rho_l = \rho_{2l} + \rho_{2l+1}, \quad \phi_{ext,l} = \phi_{ext,2l} + \phi_{ext,2l+1}, \quad \rho_{ext,l} = \rho_{ext,2l} + \rho_{ext,2l+1}.$$ The sum over $l'$ in Eq.~(\ref{27}) can be converted into an integral as $Q_l$ are densely distributed for large $L$. Identifying $(2\pi/L)\delta(Q-Q') = \delta_{l,l'}$, $Q_l$ and $Q_{l'}$ with $Q$ and $Q'$, respectively, as well as $(L/2\pi)\rho_l$ with $\rho(Q)$ so that $\rho(z) = \int^\infty_0 dQ \cos(Qz)\rho(Q)$, we find
\begin{equation}
\frac{L}{2\pi}\sum_{l'} \mathcal{M}_{ll'} \rho_{l'} = \frac{k^2 \delta\chi}{\epsilon_\perp}\frac{4\pi \rho(Q)}{k^2+Q^2} + \left(\frac{\chi_\perp}{\epsilon_\perp} - \frac{\delta\chi}{\epsilon_\perp}\frac{k^2}{k^2+Q^2}\right) 4\pi \bar{\rho}
\end{equation}
Here $$\bar{\rho} = \int^\infty_0 \frac{dQ}{\pi} \frac{k}{k^2+Q^{2}} \rho(Q)$$ is independent of $Q$. Now Eq.~(\ref{28}) can be manipulated to obtain
\begin{equation}
\rho(Q) = \frac{\left[(\epsilon_\parallel- 1) k^2 + (\epsilon_\perp - 1) Q^2  \right]\bar{\rho} + S(Q)}{\epsilon_\parallel k^2+\epsilon_\perp Q^2}, \label{30}
\end{equation}
where the source reads
\begin{equation}
S(Q) = (k^2+Q^2)\left[k^2 \delta\chi \phi_{ext}(Q) + (1- \epsilon_\perp) \rho_{ext}(Q) + \frac{2}{\pi}\chi_\perp \phi'_{ext}(0)\right]. \label{31}
\end{equation}
Here $\phi_{ext}(Q) = (L/2\pi)\phi_{ext,l}$ and $\rho_{ext}(Q) = (L/2\pi)\rho_{ext,l}$. Multiplying Eq.~(\ref{30}) by $k/(\pi(k^2+Q^2))$ and integrating it over $Q$, we arrive at
\begin{equation}
\bar{\rho} = \frac{2\epsilon}{\epsilon + 1} \int^\infty_0 \frac{dQ}{\pi}\frac{k}{k^2+Q^2}\frac{S(Q)}{\epsilon_\parallel k^2 + \epsilon_\perp Q^2} \label{32}
\end{equation}
where we have used the following integral that can be easily evaluated by the method of contour integration
\begin{equation}
\int^\infty_0 \frac{dQ}{\pi}\frac{k \cos(Qd)}{k^2+Q^2}\frac{(\epsilon_\parallel - 1)k^2 + (\epsilon_\perp - 1) Q^2}{\epsilon_\parallel k^2 + \epsilon_\perp Q^2} = \frac{1}{2}\left(e^{-kd} - e^{-kd/\gamma} \frac{1}{\epsilon}\right). \label{33}
\end{equation}
Equations (\ref{30}) - (\ref{32}) completely determine the response of a semi-infinite LHA dielectric. 

These equations can be modified to describe the responses of an infinite medium without any surfaces. In the latter case, the term with $\bar{\rho}$ -- which arises only due to surfaces -- does not exist in the expression of $\rho(Q)$, Eq.~(\ref{30}), and the term with $\phi'_{ext}(0)$ does not exist in $S(Q)$ either, Eq.~(\ref{31}). An alternative rational leading to this modification goes by assuming that the probe charges be placed far from the surfaces so that $\phi'_{ext}(z)$ vanishes near the surfaces and that one looks only at the polarization charges far from the surfaces, so that the contribution from the term with $\bar{\rho}$, which decays away from the surfaces, can be neglected. Then one finds $$\rho(Q) = - \frac{(\epsilon_\parallel - 1)k^2 + (\epsilon_\perp - 1)Q^2}{\epsilon_\parallel k^2 + \epsilon_\perp Q^2} \rho_{ext}(Q) = \left(\frac{1}{\epsilon_\parallel}\frac{k^2+Q^2}{k^2 + \gamma^2 Q^2} - 1\right)\rho_{ext}(Q)$$ for an infinite medium.

In the isotropic case with $\epsilon_\parallel = \epsilon_\perp = \epsilon$ and $\chi_\parallel = \chi_\perp = \chi$, it follows from Eq.~(\ref{30}) that $\rho(Q) = (1-\epsilon^{-1})\bar{\rho} + \tilde{S}(Q)/\epsilon$ and $$\bar{\rho} = \frac{2}{\epsilon +1} \int^\infty_0 \frac{dQ}{\pi} \frac{k}{k^2 + Q^2} \tilde{S}(Q),$$ where $$\tilde{S}(Q) = S(Q)/(k^2+Q^2) = (1-\epsilon)\rho_{ext}(Q) + 2\chi \phi'_{ext}(0)/\pi.$$ It is useful to write $\rho(Q) = \rho_1(Q) + \rho_2(Q)$, with $\rho_1(Q)$ and $\rho_{2}(Q)$ stemming from the first and the second term of $\tilde{S}(Q)$, respectively. It is easy to show that $$\rho_{2} = \frac{1}{\pi^2}\frac{\epsilon - 1}{\epsilon - 1} \phi'_{ext}(0),$$ which is the only contribution if the probe charges are located outside the film (i.e. $\rho_{ext}(Q) \equiv 0$). As $\rho_2$ is independent of $Q$, its leads to charges localized on the surface, the areal density of which can be shown to be $\frac{\phi'_{ext}(0)}{2\pi}\frac{\epsilon-1}{\epsilon + 1}$. This result is of course well known. As for $\rho_1(Q)$, it contributes to the surface charges but also volume charges, the latter amounting to $(\epsilon^{-1} - 1)\rho_{ext}(Q)$, which is nothing but the polarization charge that would exist in an infinite medium, see above. The total charges in the volume then is $\rho_{ext}/\epsilon$, as expected. 

\section{Examples}
\label{sec:4}
\subsection{A point charge outside the film}
\label{sec:4.1}
Let us consider a point probe charge of strength $\mathcal{Q}$ placed at $z = - d < 0$, a distance of $d$ exterior to the surface at $z=0$. It follows that $\rho_{ext,n} = 0$, 
\begin{equation}
\phi'_{ext}(0) = - 2\pi \mathcal{Q} e^{-kd}, \quad \phi'_{ext}(L) = - 2\pi \mathcal{Q} e^{-k(L+ d)}
\end{equation}
and
\begin{equation}
\phi_{ext,n} = \frac{2\pi}{L_n}\frac{\mathcal{Q}}{k^2+q^2_n}e^{-kd}\left(1 - (-1)^n e^{-kL}\right). \label{35}
\end{equation}
With these expressions we find
\begin{equation}
S^+_l = - \frac{\mathcal{Q}(2-\delta_{l,0})}{2L} e^{-kd}\left(1-e^{-kL}\right)\left[(\epsilon_\parallel -1)k^2 + (\epsilon_\perp -1)q^2_{2l}\right]
\end{equation}
as well as
\begin{equation}
S^-_l = - \frac{\mathcal{Q}}{L} e^{-kd}\left(1+ e^{-kL}\right)\left[(\epsilon_\parallel -1)k^2 + (\epsilon_\perp -1)q^2_{2l+1}\right],
\end{equation}
which can be inserted in Eqs.~(\ref{20}) -  (\ref{200}) to obtain 
\begin{equation}
\rho^+_l = \frac{1}{1-\Lambda_+}\frac{S^+_l}{\epsilon_\parallel k^2 + \epsilon_\perp q^2_{2l}}, ~~ \Lambda_+ = \frac{1-e^{-kL}}{L}\sum_l \frac{k(2-\delta_{l,0})}{k^2+q^2_{2l}}\frac{(\epsilon_\parallel -1)k^2 + (\epsilon_\perp -1)q^2_{2l}}{\epsilon_\parallel k^2 + \epsilon_\perp q^2_{2l}}.
\end{equation}
and analogously
\begin{equation}
\rho^-_l = \frac{1}{1-\Lambda_-}\frac{S^-_l}{\epsilon_\parallel k^2 + \epsilon_\perp q^2_{2l+1}}, ~~ \Lambda_- = \frac{1+ e^{-kL}}{L}\sum_l \frac{2k}{k^2+q^2_{2l+1}}\frac{(\epsilon_\parallel -1)k^2 + (\epsilon_\perp -1)q^2_{2l+1}}{\epsilon_\parallel k^2 + \epsilon_\perp q^2_{2l+1}}.
\end{equation}
For large $L$, $\Lambda_+\approx \Lambda_-$. 

\begin{figure*}
\begin{center}
\includegraphics[width=0.95\textwidth]{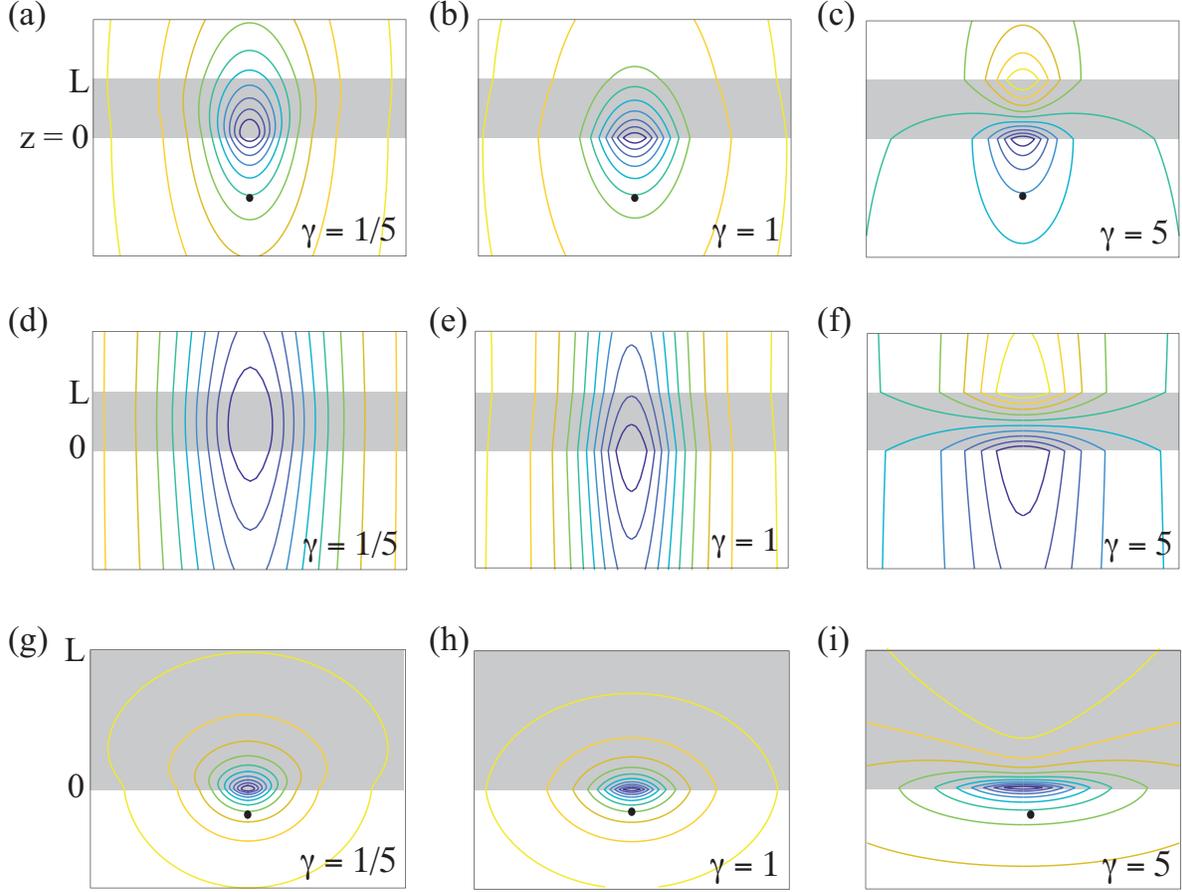}
\end{center}
\caption{Contour lines of the electrostatic potential generated by the polarization charges induced by a probe charge (indicated by a black dot) lying at $z=-d$ outside the film (shaded region) of thickness $L$. (a) - (c): d = L. (d) - (f): d = 10L. In these panels, the probe charge is not shown. (g) - (i): d = 0.1L. In these plots, $\epsilon = \sqrt{\epsilon_\parallel\epsilon_\perp} = 10$. $\gamma = \sqrt{\epsilon_\perp/\epsilon_\parallel}$ denotes the anisotropy parameter. The probe charge is not ideally point-like in these plots. Rather, it has a spread $\sim 1/k_c = 0.1d$ within the $x-y$ plane. \label{fig:2}}
\end{figure*} 

To obtain some analytical insight, let us look at the SIM limit. Equation (\ref{35}) then transforms into the following
\begin{equation}
\phi_{ext}(Q) = \frac{4\mathcal{Q}}{k^2+Q^2}e^{-kd}.
\end{equation}
Substituting this into Eq.~(\ref{31}) yields
\begin{equation}
S(Q) = - \frac{\mathcal{Q}}{\pi}e^{-kd}\left[(\epsilon_\parallel - 1)k^2 + (\epsilon_\perp - 1)Q^2\right]. \label{37}
\end{equation}
With this we get from Eqs.~(\ref{30}) and (\ref{32}) that
\begin{equation}
\bar{\rho} = - \frac{\mathcal{Q} e^{-kd}}{\pi} \frac{\epsilon - 1}{\epsilon + 1}, \quad \rho(Q) = \frac{2\epsilon}{\epsilon +1} \frac{S(Q)}{\epsilon_\parallel k^2 + \epsilon_\perp Q^2}. \label{38}
\end{equation}
In the isotropic limit this reduces to $\rho_2$ discussed in Sec.~\ref{sec:3.3}. The corresponding electrostatic potential is 
\begin{equation}
\phi(z) = \frac{2\pi}{k}e^{kz}\int^\infty_0 dQ \frac{k \rho(Q)}{k^2 + Q^2} =  \frac{2\pi}{k} \pi \bar{\rho} e^{kz}, \quad \text{for}~z<0,  \label{50}
\end{equation}
which is the same as would be produced by a fictitious point charge -- the image charge -- of value $\pi \bar{\rho} = -\mathcal{Q}(\epsilon -1)/(\epsilon +1)$ located at $z=d$. 

In Fig.~\ref{fig:2} is displayed the equipotentials of $\phi(\mathbf{x})$, i.e. the potential produced by the polarization charges only, for a variety of situations specified by $L/d$ and $\gamma$. A black dot indicates the position of the point probe charge $\mathcal{Q}$. We calculate the contour lines by $\phi(\mathbf{x}) = (1/4\pi^2)\int d^2\mathbf{k} e^{i\mathbf{k}\cdot\mathbf{r}} \phi_\mathbf{k}(z)$. Considering that $\phi_\mathbf{k}(z)$ for a point charge actually depends only on the magnitude of $\mathbf{k}$ not its direction, we can rewrite this expression as $\phi(\mathbf{x}) = (1/2\pi)\int^\infty_0 dk k J_0(kr)\phi_k(z)$, where $J_0(kr)$ is the zeroth order Bessel function of the first kind. In the numerical calculation, we have replaced upper bound of the integral by a cut-off $k_c = 10/d$. Physically, this means the probe charge is smeared over an area $\sim 1/k^2_c$ rather than being ideally point-like, i.e. $\rho_{ext}(\mathbf{x}) = \mathcal{Q}\delta(z)(1/4\pi^2)\int d^2\mathbf{k} e^{i\mathbf{k}\cdot\mathbf{r}} \rightarrow \mathcal{Q}\delta(z) (1/2\pi) \int^{k_c}_0 dk k J_0(kr)$. Only in the limit $k_c \rightarrow \infty$ does this expression represent an ideal point charge. We observe several features worth attention. Firstly, for $\gamma \gg 1$, as seen in panels (c), (f) and (i), the polarization charges are mostly concentrated near the surfaces, regardless of $L/d$. For larger $L/d$, these charges appear on both surfaces, while for smaller $L/d$, they are more concentrated on the surface closer to the probe charge. Secondly, for $\gamma \ll 1$, a significant amount of volume charges appear though only if $L/d<1$, as seen in panels (a), (d) and (g). Thirdly, in the isotropic case, $\gamma = 1$, polarization charges are concentrated on the surface closer to the probe charge regardless of $L/d$.  

\subsection{A point charge inside the film}
\label{sec:4.2}
Now let us suppose that the point probe charge is located at $z = d \in (0,L)$ inside the film, for which $\rho_{ext,n} = (\mathcal{Q}/L_n) \cos(q_nd)$ and
\begin{equation}
\phi'_{ext}(0) = 2\pi \mathcal{Q} e^{-kd}, \quad \phi'_{ext}(L) = - 2\pi \mathcal{Q} e^{-k(L-d)},
\end{equation}
as well as
\begin{equation}
\phi_{ext,n} = \frac{2\pi}{L_n}\frac{\mathcal{Q}}{k^2+q^2_n} \left(2\cos(q_nd) - e^{-kd} - (-1)^n e^{-k(L-d)}\right). 
\end{equation}
It follows that
\begin{equation}
S^+_l = \frac{\mathcal{Q}(2-\delta_{l,0})}{2L}\left(2\cos(q_{2l}d) - e^{-kd} - e^{-k(L-d)}\right)\left[(\epsilon_\parallel-1)k^2 + (\epsilon_\perp - 1)q^2_{2l}\right]. 
\end{equation}
and 
\begin{equation}
S^-_l = \frac{\mathcal{Q}}{L}\left(2\cos(q_{2l+1}d) - e^{-kd} + e^{-k(L-d)}\right)\left[(\epsilon_\parallel-1)k^2 + (\epsilon_\perp - 1)q^2_{2l+1}\right]. 
\end{equation}
These sources are then used to to obtain $\rho^\pm_l$ via Eqs.~(\ref{20}) -  (\ref{200}). In contrast with the case of the probe charge being outside, here $\rho^\pm_l$ are not simply proportional to $S^\pm_l$ due to the cosine terms in the latter. 

\begin{figure*}
\begin{center}
\includegraphics[width=0.95\textwidth]{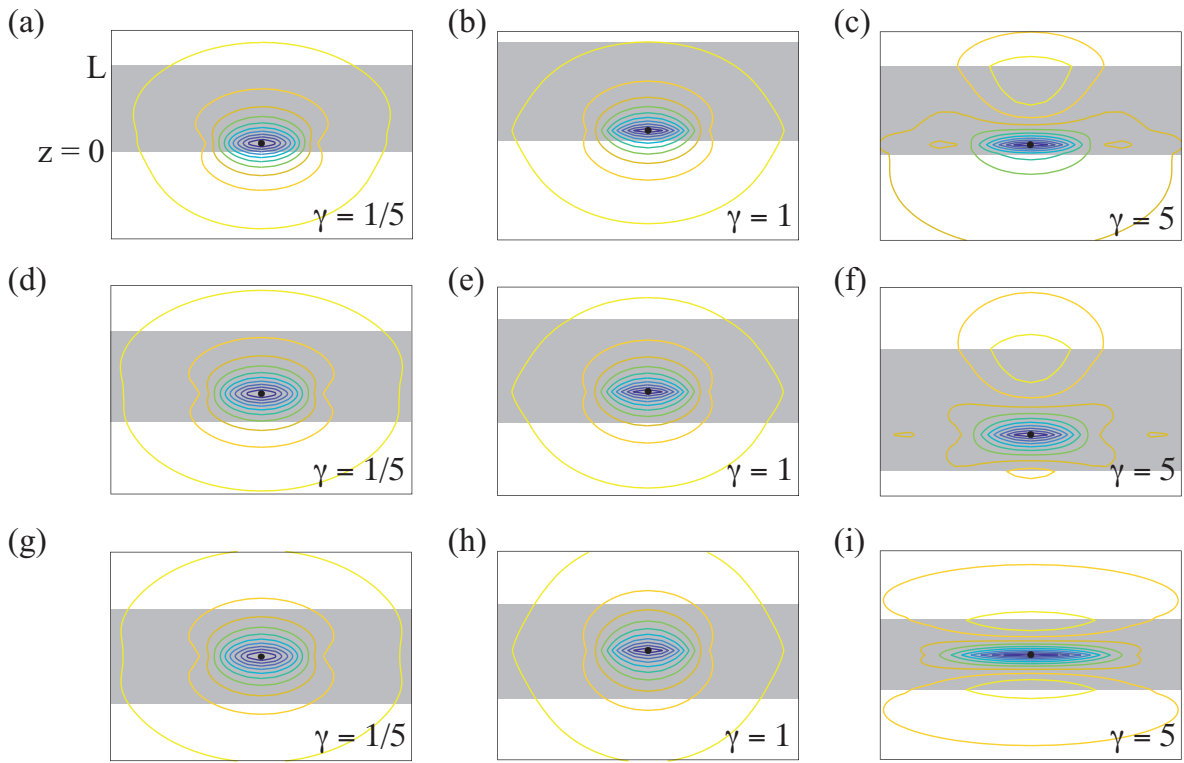}
\end{center}
\caption{Contour lines of the electrostatic potential generated by the polarization charges induced by a probe charge (indicated by a black dot) lying at $z=d$ inside the film (shaded region) of thickness $L$. (a) - (c): d = 0.1L. (d) - (f): d = 0.3L. (g) - (i): d = 0.5L. In these plots, $\epsilon = \sqrt{\epsilon_\parallel\epsilon_\perp} = 10$. $\gamma = \sqrt{\epsilon_\perp/\epsilon_\parallel}$ denotes the anisotropy parameter. As in Fig.~\ref{fig:2}, the probe charge is not ideally point-like in these plots but with a spread $\sim 1/k_c = 0.1d$ within the $x-y$ plane.\label{fig:3}}
\end{figure*} 

Again we consider the SIM limit to obtain some analytical results. Now 
\begin{equation}
\rho_{ext}(Q) = \frac{2\mathcal{Q}}{\pi}\cos(Qd), \quad \phi_{ext}(Q) = \frac{4\mathcal{Q}}{k^2+Q^2}\left(2\cos (Qd) - e^{-kd}\right). 
\end{equation}
With this we obtain
\begin{equation}
S(Q) = - \frac{\mathcal{Q}}{\pi} \left(2\cos(Qd) - e^{-kd}\right) \left[(\epsilon_\parallel - 1)k^2 + (\epsilon_\perp - 1)Q^2\right]. 
\end{equation}
This expression is very similar to the source in the case of the probe charge outside film, Eq.~(\ref{37}), but with the crucial difference of its $Q$ dependence via the factor $\cos(Qd)$. Now we find
\begin{equation}
\rho(Q) = \left(\bar{\rho} - \frac{\mathcal{Q}}{\pi}\left(2\cos(Qd) - e^{-kd}\right)\right) \frac{(\epsilon_\parallel - 1)k^2 + (\epsilon_\perp - 1)Q^2}{\epsilon_\parallel k^2 + \epsilon_\perp Q^2}
\end{equation}
together with
\begin{equation}
\bar{\rho} = - \frac{\mathcal{Q}}{\pi}\frac{1}{\epsilon+1}\left[(\epsilon + 1)e^{-kd} - 2e^{-kd/\gamma}\right].
\end{equation}
The electrostatic potential generated by $\rho(Q)$ can again be written as
\begin{equation}
\phi(z) = \frac{2\pi}{k} \pi \bar{\rho} e^{kz}, \quad \text{for}~z<0,
\end{equation}
which is of the same form as Eq.~(\ref{50}). This potential could be produced by two fictitious point charges located beneath the surface. As $\pi \bar{\rho}$ can be rewritten as $\pi \bar{\rho} = -e^{-kd}\mathcal{Q} + e^{-kd/\gamma} 2\mathcal{Q}/(1+\epsilon)$, one may say that the first fictitious charge is located at $z=d$ of strength $-\mathcal{Q}$ while the other at $z=d/\gamma$ of strength $2\mathcal{Q}/(\epsilon +1)$. It is interesting to note that, outside the film the first charge exactly cancels the potential generated by the probe charge. 

In Fig.~\ref{fig:3} we illustrate the electrostatic potential $\phi(\mathbf{x})$ generated by the polarization charges. As in Fig.~\ref{fig:2}, the probe charge, located at $z=d$ inside the film and indicated by a black dot, is not ideally point-like but smeared over an area of $\sim 1/k^2_c$ with $k_c=10/d$. In comparison with the case of an exterior probe charge (Fig.~\ref{fig:2}), here the polarization charges are all concentrated where the probe charge is located regardless of $L/d$ as long as $\gamma\leq1$. For $\gamma \geq 1$, two additional features are observed. Firstly, volume charges appear also elsewhere not just about the probe charge, as inferred from the contour lines not enclosing the probe charge. Secondly, charges also appear on the surfaces.  
  
\subsection{Plasma waves in graphene}
\label{sec:4.3}
The last example discussed here concerns plasma waves in graphene -- an atomically thin sheet of carbon atoms sitting on a honeycomb lattice. Plasma waves, which are propagating electron density ripples, in this quasi-two--dimensional world has caused a big stir in the filed of plasmonics.~\cite{GrPW} As a result of the low dimensionality, the properties of these waves are highly liable to its dielectric environment. 

\begin{figure*}
\begin{center}
\includegraphics[width=0.95\textwidth]{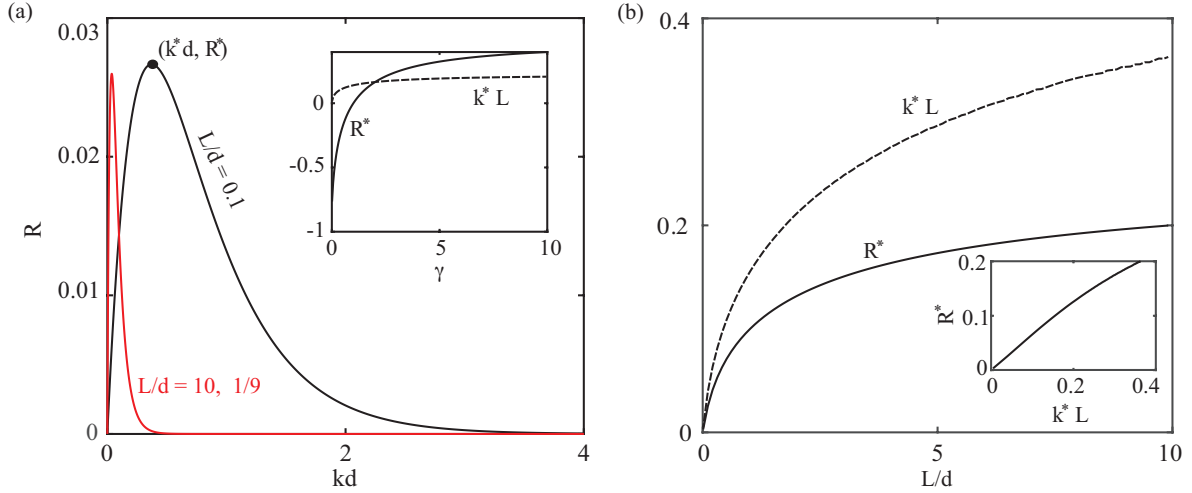}
\end{center}
\caption{Effects of an anisotropic dielectric film with constants $(\epsilon_{\parallel},\epsilon_\perp)$ of thickness $L$ on the plasma waves in a graphene sheet placed a distance $d$ off the film. $R$, which measures the effects of the anisotropy, exhibits a peak at $k^*$ with magnitude $R^*$. (a): $R$ versus the wave number $k$. The line with $L/d = 10$ has been divided by $9$ for comparison. In these plots, $\epsilon = 10, \gamma = 1.5$. Inset: anisotropy $\gamma = \sqrt{\epsilon_\perp/\epsilon_\parallel}$ dependence of $R^*$ and $k^*$ for $L/d = 0.1$. (b): thickness $L$ dependence of $R^*$ and $k^*$. In these plots, $\epsilon =10$ and $\gamma = 1.5$. Inset shows that $R^*$ is approximately a linear function of $k^*L$, i.e. $R^* \approx \alpha k^*L$, where $\alpha$ depends on $\gamma$. \label{fig:4}}
\end{figure*} 

Let us consider the simplest case of a plane charge density wave $C(\mathbf{r},t) = C e^{i(\mathbf{k}\cdot\mathbf{r} - \omega t)}$ traveling within a graphene layer placed at $z = - d$, a distance of $d$ over a LHA dielectric film. Here $C(\mathbf{r},t)$, which plays the role of $\rho_{ext}$, denotes the charge density at point $\mathbf{r}$ in the layer and moment $t$, while $\mathbf{k}$ and $\omega$ are the wave vector and frequency of the wave, respectively. The latter obey a dispersion relation $\omega(\mathbf{k})$ governed by an equation that can be shown of the following form in the electrostatic limit if the electronic collisions in the graphene layer can be neglected,~\cite{deng2015}
\begin{equation}
\omega C + ik^2\sigma \Phi_g = 0, \label{44}
\end{equation}
where $\sigma$ -- which might depend on $(\mathbf{k},\omega)$ -- is the bare electrical conductivity of the graphene layer and $\Phi_g$ is the electrostatic potential at this layer. Now $\Phi_g$ contains two contributions, one from the charges of density $\rho_{ext}(z) = C \delta(z+d)$ carried by the plasma waves and the other from the polarization charges of density $\rho(z)$ induced by $\rho_{ext}(z)$ in the dielectric underneath. The former is simply $2\pi C/k$. The latter can be written as $- (2\pi C/k) e^{-2kd} \xi,$ where
\begin{equation}
 \xi = \frac{1- e^{-kL}}{2}\frac{\Lambda_+}{1-\Lambda_+} + \frac{1+ e^{-kL}}{2}\frac{\Lambda_-}{1-\Lambda_-}.
\end{equation}
does not depend on $d$. Combined, they yield
\begin{equation}
\Phi_g = \frac{2\pi C}{k}\frac{1}{\epsilon_g}, \quad \epsilon_g = \frac{1}{1 - e^{-2kd}\xi}.
\end{equation}
We may interpret $\epsilon_g$ as the effective dielectric function felt by the charges in graphene. This is plugged in Eq.~(\ref{44}) to produce
\begin{equation}
\omega + 2\pi i k \tilde{\sigma} = 0, \quad \tilde{\sigma} = \sigma/\epsilon_g, \label{gr}
\end{equation}
which determines the dispersion of the plasma waves. The frequency is reduced in the long wavelength limit ($kd \sim 0$) but not much altered in the short wavelength limit ($kd\gg 1$). 

In the limit $kL\gg1$, one finds $\xi = (\epsilon-1)/(\epsilon +1)$, implying that a semi-infinite LHA medium effects as a LHI medium with dielectric constant $\epsilon$, as expected from Eq.~(\ref{38}). In the limit $kL\ll1$, as shown in Sec.~\ref{sec:3.2}, volume charges can be neglected in the dielectric. From the results established in Sec.~\ref{sec:3.2}, it follows that
$\chi_g = 2\beta kL/(1-\beta^2).$ This expression shows that the correction due to a thin dielectric is typically very small unless $\beta$ is close to unity, in which case the effective conductivity $\tilde{\sigma}$ might be made to change sign and no plasma waves would then exist (i.e. $\omega$ would become imaginary). The wave frequency would be widely tunable in such case by varying $d$. 

For simplicity, let us use a Drude model for graphene conductivity and write $\sigma = iF/\omega$, where $F$ is a real-valued parameter that can be tuned by doping. Then, one finds $\omega = \bar{\omega}\sqrt{kd (\tilde{\sigma}/\sigma)}$ with $\bar{\omega} = \sqrt{2\pi F/d}$. We wish to elucidate the effects of anisotropy on $\omega$ as a function of $k$. To this end, we calculate the ratio $R(k) = (\omega^2 - \omega^2_0)/\omega^2_0$, where $\omega_0(k)$ denotes the frequency had the dielectric been an LHI (also of thickness $L$) with the dielectric constant $\epsilon = \sqrt{\epsilon_\parallel \epsilon_\perp}$. The results are displayed in Fig.~\ref{fig:4} (a) for two values of $L/d = 0.1, 10$, where we see that $R$ reaches a peak $R^*$ at $k=k^*$. The inset in panel (a) shows that, while $R^*$ increases with increasing $\gamma$, $k^*$ only weakly depends on $\gamma$. On the other hand, as shown in Fig.~\ref{fig:4} (b), both $R^*$ and $k^*$ strongly depend on $L$. The former increases while the latter decreases as $L$ increases. Further, as shown in the inset of (b), we note that $R^* \approx \alpha k^*L$, where $\alpha$ is a function of $\gamma$. 

The above results may be used in the measurement of the anisotropy parameter $\gamma$ of a LHA. One may first measure out the dispersion $\omega(k)$ covering both the regimes of $kL\gg1$ and $kL\ll1$. Then $\epsilon$ can be inferred from this measured $\omega(k)$ in the limit $kL\gg1$. With $\epsilon$, $\omega_0$ can then be computed, which can be used to obtain $R$. This method can be very accurate, because $R^*$ can be very large depending on $L$. For the example shown in Fig.~\ref{fig:4} (b), $R^*$ reaches $20\%$ for $L\sim 5d$ and can be increased further with thicker films. 

\section{Conclusions}
\label{sec:5}
To summarize, we have presented a formalism for calculating the electrostatic responses, i.e. retardation effects excluded, of a LHA film with a diagonal but anisotropic susceptibility tensor. Generalization to an arbitrary tensor is straightforward but not included in this work. The most interesting aspect of our derivation is that, a generic macroscopic physical description of surfaces is prescribed, which does away with the boundary conditions that are essential in traditional textbook approaches. Our formalism makes it clear that the responses consist of two contributions, one stemming from the very presence of surfaces while the other existing even in an infinite medium. We have illustrated the formalism with three examples. In one of these we discuss the plasma waves in graphene under the influence of a LHA film. It is shown that the frequency of the waves is strongly affected at wavelength comparable to the film thickness. 

While we explicitly speak of a dielectric, our formalism can be equally applied to calculate the electrical responses (e.g. current flow) in an anisotropic conductor by the duality relation that exists between conductivity tensor and susceptibility tensor.

\end{document}